# Single-axis dependent structural and multiferroic properties of orthorhombic $R$MnO$_3$ ($R$ = Gd – Lu)

Kenta Shimamoto,[1] Saumya Mukherjee,[2] Nicholas S. Bingham,[1,3,†] Anna K. Suszka,[1,3] Thomas Lippert,[1,4] Christof Niedermayer,[2] Christof W. Schneider,[1,*]

[1] *Laboratory for Multiscale Materials Experiments, Paul Scherrer Institut, CH 5232 Villigen-PSI, Switzerland*

[2] *Laboratory for Neutron Scattering and Imaging, Paul Scherrer Institut, CH 5232 Villigen-PSI, Switzerland*

[3] *Laboratory of Mesoscopic Systems, Department of Materials, ETH Zürich, CH-8093 Zurich, Switzerland*

[4] *Laboratory of Inorganic Chemistry, Department of Chemistry and Applied Biosciences, ETH Zürich, CH 8093 Zurich, Switzerland*

[†] Present affiliation: U.S. Naval Research Laboratory, 20375 Washington D.C., USA

[*]Correspondence and requests for materials should be addressed to C.W.S. (email: christof.schneider@psi.ch).

**Abstract**

Controlling material properties by modulating the crystalline structure has been attempted using various techniques, e.g., hydrostatic pressure, chemical pressure, and epitaxy. These techniques succeed to improve properties and achieve desired functionalities by changing the unit cell in all dimensions. In order to obtain a more detailed understanding on the relation between the crystal lattice and material properties, it is desirable to investigate the influence of a smaller number of parameters. Here, we utilize the combination of chemical pressure and epitaxy to modify a single lattice parameter of the multiferroic orthorhombic $R$MnO$_3$ ($R$ = rare-earth; $o$-$R$MnO$_3$) system. By growing a series of $o$-$R$MnO$_3$ ($R$ = Gd – Lu) films coherently on (010)-oriented YAlO$_3$ substrates, the influence of chemical pressure is reflected only along the $b$-axis. Thus, a series of $o$-$R$MnO$_3$ with $a \sim 5.18$ Å, $5.77$ Å $< b < 5.98$ Å, and $c \sim 7.37$ Å were obtained. Raman spectra analysis reveals that the change of the $b$-axis parameter induces a shift of the oxygen in the nominally "fixed" $ca$-plane. Their ferroelectric ground state is independent on the $b$-axis parameter showing polarization of ~ 1 µC cm$^{-2}$ along the $a$-axis for the above-mentioned range, except for $b \sim 5.94$ Å which corresponds to TbMnO$_3$ showing ~ 2 µC cm$^{-2}$. This result implies that multiferroic order of $o$-$R$MnO$_3$ is almost robust against the $b$-axis parameter provided that the dimension of the $ca$-plane is fixed to 7.37 Å × 5.18 Å.

## I. INTRODUCTION

To tailor functionalities and to find new properties in materials, the artificial tuning of lattice parameters has been used as a basic but effective approach. The lattice parameters are directly linked to the bond lengths and angles in a crystal, thereby governing the electronic distribution and the energy landscape of material systems. Mechanical stress such as hydrostatic pressure is a straightforward method and has been widely utilized to examine the lattice-properties relationship of materials [1]. The applicable pressure range has been widened by the advancement of pressure cells such as diamond anvil cells [2,3], inducing progress in the search for exotic phases and functionalities in materials [4-11]. Another well-known approach is the substitution of selected atoms by atoms with a different radius and the same valence state, which is called chemical pressure. It should be noted that both occupied and unoccupied orbitals of the substituting substituted atoms do not directly contribute to the band structure near the Fermi level when chemical pressure is discussed. By changing the size of the atoms, the interatomic distances and/or the bond angles are efficiently altered [12-14], and so are the material properties [13,15-19]. A typical example is the series of orthorhombic $R$MnO$_3$ ($o$-$R$MnO$_3$, $R$ = La – Lu, Y) which exhibit a perovskite structure. The size of the $R$ ion has an influence on the rotation, the tilt, and the distortion of the MnO$_6$ octahedra along with the lattice parameters [13]. Hence their electric and magnetic ground states strongly depend on the size of the $R$ ions. For $R$ = La – Gd, A-type antiferromagnetism (AFM) without ferroelectricity is reported as a ground state [20,21]. Incommensurate cycloidal AFM together with a small ferroelectric (FE) polarization ($P$; 0.06 – 0.15 µC cm$^{-2}$) along the $c$-axis ($\|c$) is a ground state for $R$ = Tb and Dy [22,23], while E-type commensurate AFM with $P\|a$ of ~ 0.5 µC cm$^{-2}$ appears for $o$-$R$MnO$_3$ with smaller $R$ ions [24-26].

Growth-induced strain using epitaxy, so-called epitaxial strain, is an up-to-date tool to apply strain in a thin film material [27,28]. At the initial stage of the growth, a material tends to have its in-plane lattice parameters matched to the substrate. Therefore, the magnitude of the strain depends on the in-plane lattice mismatch between the targeted material and substrates. By growing epitaxial films on a series of different substrates, one can gain a fundamental understanding on lattice-property relations [29-37]. The aforementioned three individual techniques to control lattice parameters, applying hydrostatic pressure, substituting atoms, and growing films on various substrates, change all crystalline directions simultaneously. For further fundamental investigations on the relation between lattice and material properties, it is more desirable to prepare samples where only a single lattice parameter is varied with the others fixed for the sake of simplicity. This article discusses the combination of chemical pressure and epitaxy in order to achieve the aforementioned aim as is implicitly applied in Refs [38] and [39]. Here, we utilize only one type of substrate, a (010)-oriented $YAlO_3$ substrate, and coherently grow a series of $o$-$R$MnO$_3$ films ($R$ = Gd – Lu) to lock the in-plane lattice parameters. Although all the lattice parameters are modified compared to bulk, the overall series of films shows the same in-plane lattice parameters with different out-of-plane lattice parameters. Hence the out-of-plane lattice parameter dependent structural and multiferroic properties can be investigated to deepen the fundamental understandings of the $o$-$R$MnO$_3$ system.

## II. EXPERIMENT

$o$-$R$MnO$_3$ ($R$ = Gd – Lu) epitaxial films (~15 nm, actual thicknesses are listed in Fig. 4) were prepared on (010)-oriented $YAlO_3$ substrates (Crystec Co., Ltd.) by pulsed laser deposition using a KrF excimer laser ($\lambda$ = 248 nm, 2 Hz) with the cut of the 5 × 10 × 0.5 mm substrates along the [100] or [001] in-plane orientation. The laser beam was focused on a sintered ceramic target (orthorhombic for $R$ = Gd – Dy, and hexagonal for the rest) with a spot size of ~ 1.2 × 1.7 mm. The laser fluence was adjusted for each $R$MnO$_3$ target as discussed in [40] in order to minimize the influence of crystallographic qualities when comparing the structural and the physical properties of a series of $o$-$R$MnO$_3$ films. The substrate was mounted in an on-axis geometry to the plasma plume with a distance of 4.1 cm from the target. The films were grown in a N$_2$O partial pressure of 0.70 mbar with the substrate temperature of 690°C maintained by a lamp heater. Lattice parameters of the grown films were analyzed by x-ray diffraction measurements using a Seifert four-circle x-ray diffractometer with a CuK$\alpha$1 monochromatic x-ray source. Raman scattering spectra were taken by a HORIBA Jobin Yvon LabRAM HR800 confocal Raman spectrometer. The spectra were probed in the backscattering geometry with the laser light polarization parallel to the $a$-axis of the (010)-oriented films. In-plane electrical characterizations were performed by patterning Au (56 nm) / Ti (4 nm) interdigitated electrodes on the film surfaces as depicted in Ref. [41]. Measurements were performed in a continuous helium flow atmosphere and the temperature was controlled by a LakeShore Model 325 temperature controller. Capacitances were investigated using an Agilent E4980A LCR meter at frequencies between 100 Hz and 2 MHz. Data taken at 15 kHz are shown in Fig. 4(b) and the rest are shown in Ref. [40]. Ferroelectric hysteresis curves were probed through the Positive-Up Negative-Down (double-wave) method [42]. The polarization was calculated as $P = Q(tL)^{-1}$, where $Q$ is the measured charge, $t$ is the film thickness, and $L$ is the total length of the finger pairs of the interdigitated electrodes [43,44]. Further details on the electrical characterization techniques are described elsewhere [40,45]. Temperature dependent magnetization was investigated by a commercially available SQUID magnetometer (Quantum Design, MPMS® 3) with the magnetic field aligned along the long substrate axis.

Although we previously reported the structural and magnetic order of $o$-HoMnO$_3$ [41], $o$-TmMnO$_3$ [46], and $o$-LuMnO$_3$ [47] films on (010)-oriented YAlO$_3$ substrates, we do not compare their results to those presented here. The samples presented here are prepared with different growth conditions as a result of a change of a heater with a different geometry and a heat source [40,48], giving rise to films with modified and improved crystallographic qualities. The chemical and crystallographic features behind the difference in qualities from previous works are at present not well understood. The origins are most likely related to the different heater which results in an improved homogeneity of the heating.

## III. RESULTS

$R$MnO$_3$ ($R$ = Gd – Lu) are grown epitaxially on YAlO$_3$ (010) substrates as an orthorhombic phase. For $R$ = Ho – Lu, where the stable crystalline phase is hexagonal [49], the perovskite substrate with small lattice mismatches stabilizes the metastable orthorhombic phase similar to previous reports on the film growth [50,51]. If the grown films are thin enough, their in-plane lattice parameters can be locked to those of the substrate as illustrated by the reciprocal space maps of the (130) and the (041) reflections [Figs. 1(a) and 1(b)]. Accordingly, the lattice parameters of the $o$-$R$MnO$_3$ films are different from those of bulk $o$-$R$MnO$_3$. All the films are compressed along the $a$-axis while the strain along the $c$-axis changes by the size of the $R$ ions. The films are also compressed along the $c$-axis for $R$ = Gd – Dy, and the $c$-axis parameter of the films is expanded for the rest of the smaller $R$ ions. Accordingly, the $o$-$R$MnO$_3$ films expand along the $b$-axis except for $o$-YbMnO$_3$ and $o$-LuMnO$_3$ films. As a consequence of the interplay between tensile and compressive strain along the different crystalline directions and the elastic properties of the material, the volume of the unit cell $V$ follows the opposite trend of the $b$-axis. When the $b$-axis is expanded by epitaxial strain from a YAlO$_3$ (010) substrate, $V$ of the film is smaller compared to bulk. Here it should be noted that all the lattice parameters of bulk materials are monotonically dependent on the size of $R$ ions, while for the coherently grown films the influence of the $R$-ion size is reflected only in the $b$-axis parameter [Fig. 1(c)]. Hence, it is possible to discuss the influence of a single lattice parameter on the properties of a material system by growing a series of $o$-$R$MnO$_3$ films coherently on the same type of substrate. The change of a single lattice parameter using epitaxy on the same substrate can also be achieved by modulating the chemical composition of the selected material. In such cases, however, chemical and crystalline imperfections in the films govern the modulated physical properties [52-54]. The use of chemical pressure allows investigating single-axis dependent properties almost without taking into account the influence of the imperfections by carefully preparing the series of samples.

For some $o$-$R$MnO$_3$ films, the magnitude of the applied strain can be deduced by comparing the lattice parameters of the films with those of bulk polycrystalline samples under hydrostatic pressure [55-57] (Fig. 2). A linear extrapolation of data from Ref. [56] shows that the compressively strained TbMnO$_3$ film experiences equivalent pressures of ~ 11 GPa ||$a$ and ~ 4 GPa ||$c$. Likewise, the strained GdMnO$_3$, DyMnO$_3$, and HoMnO$_3$ film experiences pressures of ~ 13 GPa ||$a$ and ~ 8 GPa ||$c$, ~ 9 GPa ||$a$ and ~ 0 GPa ||$c$, and ~ 7 GPa ||$a$ and ~ 0 GPa ||$c$ respectively. Accordingly, the out-of-plane parameter ($b$) is expanded by ~ 1.9 % for GdMnO$_3$, ~ 1.7 % for TbMnO$_3$, ~ 1.2 % for DyMnO$_3$, and ~ 0.7 % for HoMnO$_3$.

Figure 3(a) shows Raman scattering spectra of the series of $o$-$R$MnO$_3$ films and a single crystal YAlO$_3$ as reference. Due to the measurement geometry, only $A_g$ modes can be probed. As discussed in bulk perovskite materials, some of the Raman modes reflect the rotation angles or the distortion of BO$_6$ octahedra [58-60]. Here the following three modes, $A_g$(4), $A_g$(1), and $A_g$(3), are discussed, which correspond to the rotation of MnO$_6$ octahedra around the *b*-axis, the Jahn-Teller stretching, and the bending of MnO$_6$ octahedra, respectively [61]. First, the effect of epitaxial strain to Raman modes is discussed. Bulk Raman spectra are available for *R* = La – Ho [58], and their peak positions are shown in Fig. 3(b). All three modes are hardened by epitaxial strain (i.e., shift to larger wave numbers compared to the bulk values). As the magnitude of epitaxial strain gets larger (i.e., for larger *R* ions), the frequency shift of the Raman modes increases. This represents the fact that the vibrations of MnO$_6$ octahedra are more strongly disturbed by applying larger epitaxial strain. In other words, the Jahn-Teller distortion and the rotation around the *b*-axis of MnO$_6$ octahedra become larger by epitaxial strain. It is likely that larger compressive strain in the *ca*-plane may confine the motions of atoms more. Stronger Jahn-Teller distortion for larger *R* ions which can be roughly expected from a higher *b*/*a* ratio of the unit cell [Fig. 1(c)] is now confirmed by the Raman spectra. We also note another influence of epitaxial strain to the Raman spectra. In bulk *R*MnO$_3$, the mixing of $A_g(1)$ and $A_g(3)$ is reported especially for GdMnO$_3$ and TbMnO$_3$ [58]. The closeness of the $A_g(1)$ and the $A_g(3)$ mode frequency induces mode repulsion and intensity transfer. However, in the case of strained films, the peak of the $A_g(3)$ mode is largely isolated from the $A_g(1)$ mode (e.g. the difference is ~ 20 cm$^{-1}$ for bulk GdMnO$_3$ and ~ 30 cm$^{-1}$ for the strained GdMnO$_3$ film), therefore those two phonons are considered to be less mixed than for bulk. The intensity ratio of these two mixing modes [$A_g(1)$ vs. $A_g(3)$] is also larger for films (e.g., for GdMnO$_3$, 1.36:1 for bulk [58] and 1.66:1 for film) which further supports the separation of phonon modes by epitaxial strain.

The *R* ion dependence of the film Raman spectra brings a different perspective in the Raman mode analysis, i.e., the *b*-axis parameter dependence. Each measured mode depends on the size of the *R* ion, exhibiting a higher mode frequency for smaller *R* ions. The Jahn-Teller distortion and rotation of the MnO$_6$ octahedra increases as the *b*-axis parameter becomes smaller. The detailed analysis on the $A_g(4)$ mode gives an important insight. The $A_g(4)$ mode represents the rotation angle of MnO$_6$ octahedra around the *b*-axis. Therefore, changing the *b*-axis parameter has an influence on the rotation angle of MnO$_6$ octahedra around the *b*-axis. Here, it should be noted that the in-plane lattice parameters (the *c*- and the *a*-axis parameters) are fixed by epitaxial strain as shown in Figs. 1(c) and 1(d). This concludes that even if in-plane lattice parameters are locked by epitaxy, chemical pressure by the *R* ion has an influence on the locations of oxygen atoms along the in-plane directions. In other word, the modulation of the *b*-axis parameter induces a shift of the oxygen atoms in the "fixed" *ca*-plane.

Figure 4 shows the magnetic and electric characterization of *o*-*R*MnO$_3$ films. Magnetic transition temperatures are analyzed by temperature dependent magnetization measurements [Fig. 4(a)]. Zero-field-cooled measurements along the *c*-axis show a small hump for the GdMnO$_3$, the TbMnO$_3$, and the *o*-HoMnO$_3$ films. Those humps indicate the phase transition from paramagnetism to AFM, i.e., the Néel temperatures ($T_N$). The temperature-dependent magnetizations of the other *o*-*R*MnO$_3$ films measured along the *a*-axis do not show any humps. Instead, $T_N$ was analyzed by comparing zero-field-cooled and field-cooled measurements, where these two measurements exhibit different results due to the order of Mn spins. The separation of zero-field-cooled and field-cooled magnetization measurements as illustrated in Fig. 4(a) is often seen for *o*-*R*MnO$_3$ films [62-65]. For all the investigated *o*-*R*MnO$_3$ films, the $T_N$ are 42 – 43 K regardless of the *R* ions [Fig. 4(a), Table I] and almost correspond to bulk *o*-*R*MnO$_3$ [20,23,25]. In all the magnetization curves shown in Fig. 4(a), potential signs of a further phase transition were not observed below $T_N$ down to 10 K, similar to bulk *o*-*R*MnO$_3$ with *R* = Er – Lu [24,25,66,67]. It should be noted that the transition from incommensurate to E-type AFM is almost impossible to observe by magnetization measurements. For bulk GdMnO$_3$ and TbMnO$_3$, the second magnetic transition can be observed from magnetization curves (GdMnO$_3$: ~ 24 K [20,23], TbMnO$_3$: ~ 27 K [20,22]), which is not observed in the strained films. These results imply that the magnetic phases of the strained GdMnO$_3$ films can be different, as observed for the strained TbMnO$_3$ films [45].

The FE transition temperature ($T_{FE}$) determined by the capacitance measurements, on the other hand, depends on the *R* ion. The temperature-dependent normalized capacitance measured along the *a*-axis shows a clear divergent behavior for all the *o-R*MnO$_3$ films indicating a FE phase transition [Fig. 4(b), Table I]. No other anomaly was observed for all the films in the measured temperature range (from 8 to 50 K). The temperature-dependent remanent polarizations derived from the FE hysteresis curves demonstrate that for all the films the direction of *P* below $T_{FE}$ is along the *a*-axis [Fig. 4(c)], (See also the Supplemental Materials [68]), i.e., no other apparent FE phase transition was observed between 8 K and the $T_{FE}$. The drop of the remanent polarization at low temperatures is due to the instrumental limit of available input voltage which is not large enough to fully polarize the sample. The electric properties of the GdMnO$_3$, the TbMnO$_3$ [45,54], and the DyMnO$_3$ films differ significantly from bulk. All exhibit higher $T_{FE}$ compared to bulk (Table I), the direction of *P* changes from the *c*- to the *a*-axis for TbMnO$_3$ and DyMnO$_3$, and the magnitude of *P* is enhanced by more than ten times [23]. Since *o-R*MnO$_3$ exhibits spin-driven ferroelectricity [69], these changes of electric properties also imply that their magnetic states are significantly altered. The direction and the magnitude of *P* shown in Figs. 4(c) and 4(d) correspond to the FE state induced by E-type AFM as observed and discussed in the case of bulk TbMnO$_3$ under hydrostatic pressure [8,70] and strained films [45]. The change of FE properties in the strained GdMnO$_3$ and DyMnO$_3$ films observed here also resembles bulk under hydrostatic pressure [71] except for the $T_{FE}$, which is higher for the films strained by the (010)-oriented YAlO$_3$ substrates (Table I). The magnitude of $P\|a$ for the rest of the $o$-$R$MnO$_3$ films is ~ 1 µC cm$^{-2}$ [Fig. 4(c)] which is about two times larger than the values for the bulk materials [25]. Compared to bulk results, $o$-HoMnO$_3$, $o$-ErMnO$_3$, and $o$-TmMnO$_3$ films exhibit a higher $T_{FE}$ while $o$-YbMnO$_3$ and $o$-LuMnO$_3$ films show lower values as a consequence of epitaxial strain (Table I). The existence of the $P\|a$ ~ 1 µC cm$^{-2}$ implies that the magnetic ground state of these films is E-type AFM as suggested in the case of TbMnO$_3$ films [45]. Incommensurate magnetic order is another potential candidate as reported for $o$-YMnO$_3$ films grown on the same substrate [65,72].

**IV. DISCUSSIONS**

Figure 5 compares the multiferroic phase diagram of the bulk materials and of the $o$-$R$MnO$_3$ films coherently grown on (010)-oriented YAlO$_3$ substrates. The bulk phase diagram [Fig. 5(a)] has the following two main features. (1) The diagram consists of mainly three phases depending on the size of $R$ ion, A-type AFM without $P$, $bc$-cycloidal AFM with $P\|c$, and E-type AFM with $P\|a$. (2) The $T_{FE}$ increases for $o$-$R$MnO$_3$ with smaller $R$ ions. Each feature is, as shown by Monte Carlo simulations, correlated to the change of the dominant multiferroic mechanism and the increase of next-nearest-neighbor exchange interaction between Mn along the $b$-axis ($J_b$), respectively [73,74]. These two main trends are clearly altered by epitaxial strain. From a FE point of view, the diagram has only one phase for $R$ = Gd – Lu with a $P\|a$ ~ 1 µC cm$^{-2}$ (except for $R$ = Tb with $P\|a$ ~ 2 µC cm$^{-2}$) and the trend of $T_{FE}$ against the size of $R$ ion is inverted by epitaxial strain. Considering the magnitude and the direction of $P$, the phase diagram implies that the symmetric magnetostriction contributes to the multiferroic ground state of $o$-$R$MnO$_3$ films as discussed for TbMnO$_3$ [45]. To verify this hypothesis the magnetic order of all the $o$-$R$MnO$_3$ films must be investigated.

It can be claimed that epitaxial strain rewrites the multiferroic phase diagram of $o$-$R$MnO$_3$, changing the energy landscape of the system. The newly generated phase diagram in Fig. 5(b) is attributed to the strain introduced by a YAlO$_3$ (010) substrate. In the example shown here, the role of epitaxy is limited to the lattice deformation for simplicity. However epitaxial growth is also capable of inducing defects and domain walls. By changing the substrate material and orientation, lattice parameters, microstructures, and physical properties of the material vary [75-81]; therefore a different phase diagram can be realized.

The horizontal axis of the diagram in Fig. 5(b), i.e., the ionic $R$ radius, can be replaced by the $b$-axis lattice parameter since the films are coherently grown and their $ca$-planes are locked [Fig. 1(c)]. Therefore, the diagram leads to the following conclusions. When the $ca$-plane is locked to the (010)-oriented YAlO$_3$ substrate, the $T_{FE}$ shows a positive correlation to the $b$-axis, exhibiting a smaller $T_{FE}$ for a smaller $b$-axis parameter. From the observed FE ground states of the series of films, it is expected that there is a contribution of the symmetric magnetostriction as already discussed. It is reported that the nearest-neighbor and next-nearest-neighbor exchange interactions between Mn ions in the $ab$-plane ($J_{ab}$ and $J_b$, respectively) play a crucial role for its multiferroic ground state [73,82,83]. Following the Monte Carlo simulation results in Ref. [73], it can be suggested that the $ca$-plane with 7.37 Å × 5.18 Å causes a large $J_b$ (or the ratio of $J_b$ and $J_{ab}$, $|J_b/J_{ab}|$ [82,83]) which induces a FE state originating from the symmetric magnetostriction within the presented $b$-axis range, not only in the case of TbMnO$_3$ [45]. The decrease of $T_{FE}$ caused by a smaller $b$-axis parameter in the series of $o$-$R$MnO$_3$ films [Fig. 5(b)] may be attributed to the decrease of $J_b$, judging from the same Monte Carlo simulation results.

From the Raman spectra analysis of the $A_g(4)$ mode, the potential decrease of $J_b$ by substituting a smaller $R$ ion can be estimated in the strained films. As discussed in Ref. [61], the frequency of the $A_g(4)$ mode corresponds to the rotation angle of MnO$_6$ octahedra around the *b*-axis. The higher the frequency, the more the oxygen atoms are shifted with respect to the Mn atom as illustrated in Figs. 3(c) and 3(d). This analysis agrees with the following instinctive understanding: Oxygen atoms shift to compensate for the change of the size of *R* ions. When a smaller *R* ion is substituted, oxygen atoms move towards the *R* ion to fill the space. Since the exchange interaction $J_b$ is mediated by two oxygen atoms [O(2) and O(3) in Fig. 3(d)], its magnitude depends largely on the orbital overlap between these oxygen atoms. The *bc*-cross section of the unit cell shown in Fig. 3(d) implies that the orbital overlap tends to be smaller as the $A_g(4)$ mode frequency becomes larger. Therefore the *o-R*MnO$_3$ film with a smaller *R* ion may have a lower $J_b$. Assuming that $J_{ab}$ is almost independent of the *R* ion as calculated in the case of the bulk materials for *R* = Gd – Er [84], this lowering of $J_b$ may contribute to the decrease of $T_{FE}$ for films with a smaller *R* ion as indicated in the calculated phase diagram in Refs. [73,74]. It has to be noted that the same trend can be found in the $A_g(4)$ mode of bulk samples. Hence, the discussion only of the $A_g(4)$ mode cannot explain the trend of $J_b$. It is at least required to clarify the shift of oxygen atoms in the *ab*-plane, too. The $A_g(2)$ Raman mode corresponds to MnO$_6$ rotation angle around the *c*-axis, which is useful for qualitative analysis of the oxygen atoms in the *ab*-plane. However, the intensity of the mode is more than five times weaker than the $A_g(4)$ mode and cannot be detected for the ~ 15 nm films. Furthermore, a strong peak from the YAlO$_3$ substrate at around 340 cm$^{-1}$ also disturbs the detection of the weak $A_g(2)$ mode.

In the case of multiferroic *o-R*MnO$_3$, as reported in theoretical works, it is necessary to discuss bond angles and atomic locations in order to address their physical properties. Therefore, obtaining interatomic distances and bond angles of nm-thick films is indispensable for further understandings. Such properties recently start to be investigated with annular bright-field imaging in aberration-corrected scanning transmission electron microscopy [85-87] for thin-film cross-sections. By gaining a direct correlation between single-axis parameter and atomic locations, condensed matter research with oxide thin films should make further steps forward.

## V. CONCLUSION

$o$-$R$MnO$_3$ ($R$ = Gd – Lu) films are grown coherently on (010)-oriented YAlO$_3$ substrates and their lattice and multiferroic properties are investigated. By growing a series of $o$-$R$MnO$_3$ films coherently on the same substrate, the influence of chemical pressure is reflected only in the out-of-plane ($b$-axis) lattice parameter. Raman spectra analysis reveals that the change of $R$ ion, i.e., the modulation of the $b$-axis parameter, induces a shift of oxygen atoms in the $ca$-plane. The multiferroic phase diagram of $o$-$R$MnO$_3$ has been modified by epitaxial strain, exhibiting only one ferroelectric ground state with the polarization along the $a$-axis. Interpreting the results from single-axis dependence, it can be concluded that FE transition temperature mostly shows a positive correlation to the $b$-axis parameter potentially due to the increase of exchange parameter along the $b$-axis. In the meantime, the ferroelectric ground state remains the same despite the change of the $b$-axis parameter by ~ 3.5 %, suggesting that the ground state of $o$-$R$MnO$_3$ is not sensitive against the $b$-axis parameter with the constant $ca$-plane dimension of 7.37 Å × 5.18 Å.

**Acknowledgements**

We would like to thank D. Marty (Laboratory for Micro- and Nanotechnology, PSI) for support with optical lithography and U. Greuter (Laboratory for Particle Physics, PSI) for the implementation of a Sawyer-Tower circuit for ferroelectric hysteresis measurements. The use of the LCR meter belonging to the Laboratory for Scientific Developments and Novel Materials, PSI, is gratefully appreciated. The drawings of crystal structures in Fig. 3 are produced by VESTA program [88], which is gratefully acknowledged Financial support and CROSS funding to K.S. from PSI are gratefully acknowledged. S.M. acknowledges financial support from the Swiss National Science Foundation (SNSF Project No. 200021_147049). A.K.S. acknowledges the funding from the European Community's Seventh Framework Program (FP7/2007-2013) under Grant Agreement No. 290605 (COFUND: PSI-FELLOW).

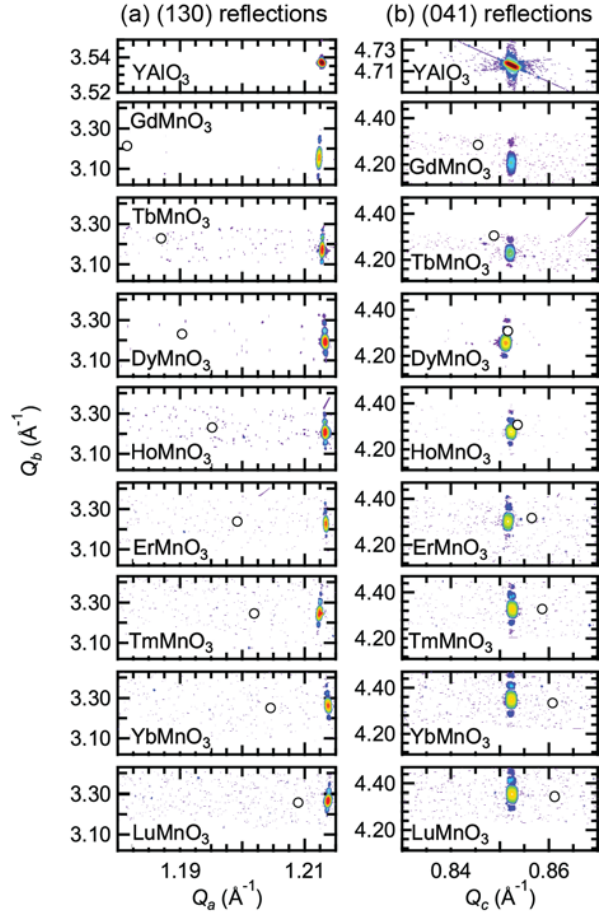

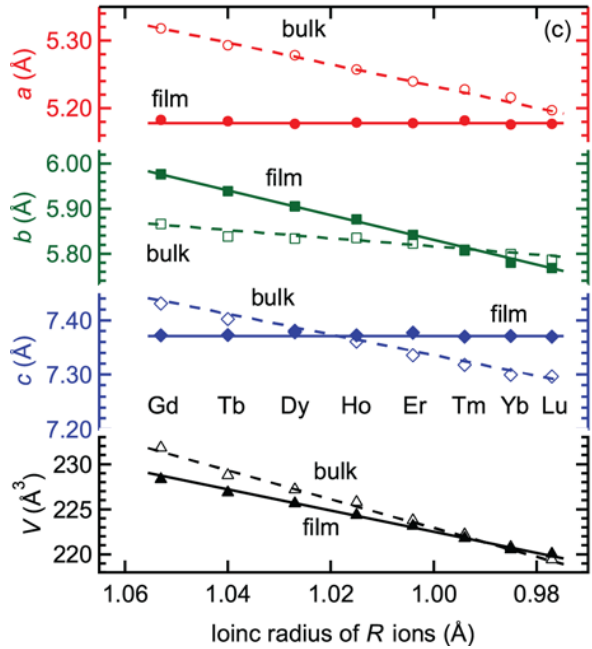

Fig. 1. Reciprocal space maps of (a) the (130) reflection and (b) the (041) reflection of $o$-$R$MnO$_3$ films ($R$ = Gd – Lu) grown on (010)-oriented YAlO$_3$ substrates. Markers on each map correspond

to the locations for bulk [13,89,90]. (c) Lattice parameters and unit cell volumes of a series of $o$-$R$MnO$_3$ ($R$ = Gd – Lu) plotted as a function of the ionic radius of $R$ ions. Open and closed symbols correspond to bulk (Gd: Ref. [90], Tb – Ho: Ref. [89], Er – Lu: Ref. [13]) and films coherently grown on (010)-oriented YAlO$_3$ substrates, respectively. Solid and dashed straight lines are guides to the eyes.

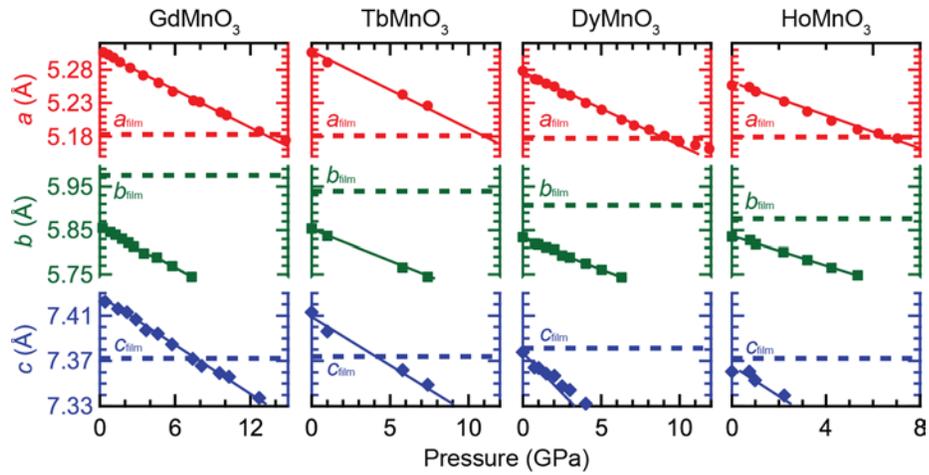

Fig. 2. Comparison of lattice parameters of the $o$-$R$MnO$_3$ films ($R$ = Gd – Ho) to those of bulk under hydrostatic pressure [55-57]. Horizontal dashed lines show the lattice parameters of thin films calculated from Figs. 1(a) and 1(b) and markers are from data in the references. Solid straight lines are guides to the eyes.

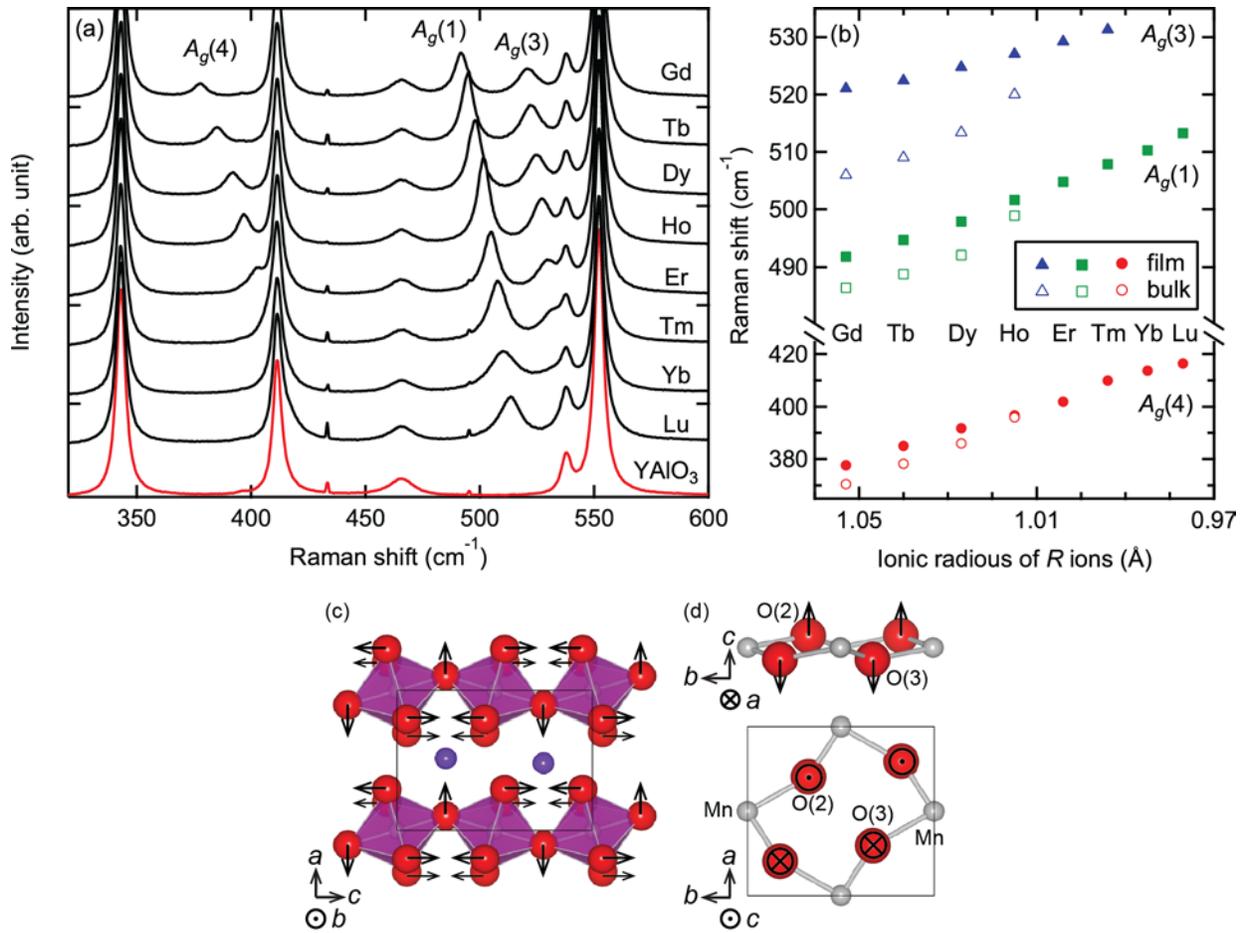

Fig. 3. (a) Raman spectra of a series of *o*-$R$MnO$_3$ films. A spectrum of a (010)-oriented YAlO$_3$ substrate is shown at the bottom. (b) $R$-ion-radius dependence of the Raman shift of $A_g$(1), $A_g$(3), and $A_g$(4) modes. Assignment of phonon modes in (a) and bulk values in (b) are adapted from Ref. [58]. The direction of the shift of oxygen atoms represented by the $A_g$(4) Raman mode frequency is depicted in (c) MnO$_6$ octahedra and (d) selected Mn and O atoms. Arrows indicate the direction of shift by increasing the mode frequency, i.e., by substituting the $R$ ion with a smaller one.

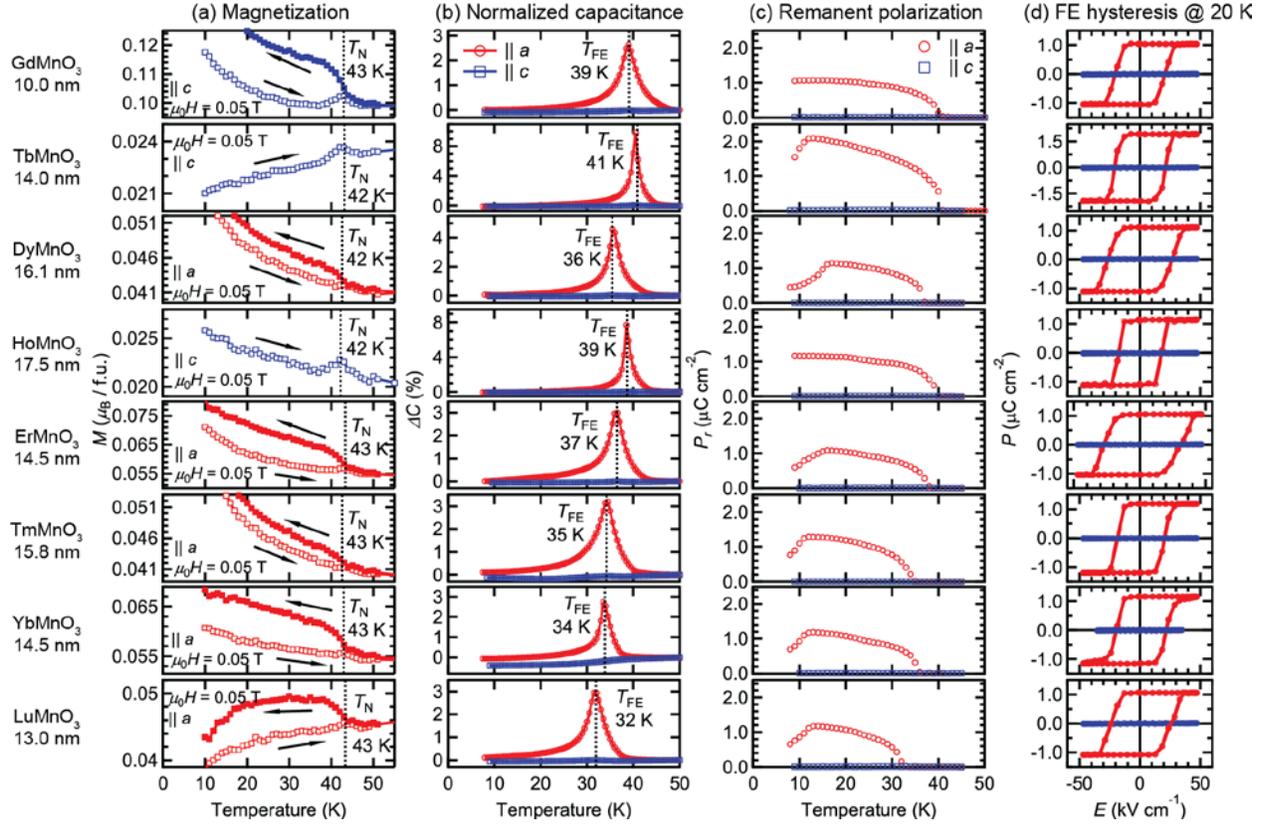

Fig. 4. Temperature dependent (a) field-cooled (closed symbols) and zero-field-cooled (open symbols) magnetization measurements at $\mu_0 H = 0.05$ T, (b) normalized capacitance ($\Delta C = [C(T) - C(50\ \text{K})]/C(50\ \text{K})$), (c) remanent polarization, and (d) FE hysteresis curves of $o$-$R$MnO$_3$ films coherently grown on (010)-oriented YAlO$_3$ substrates. The drop in remanent polarization at low temperatures is due to the instrumental limit of available input voltage which is not large enough to fully polarize the sample. As an example of how the polarization measurements were conducted, raw data of the FE hysteresis curve of TbMnO$_3$ are shown in the Supplemental Material [68].

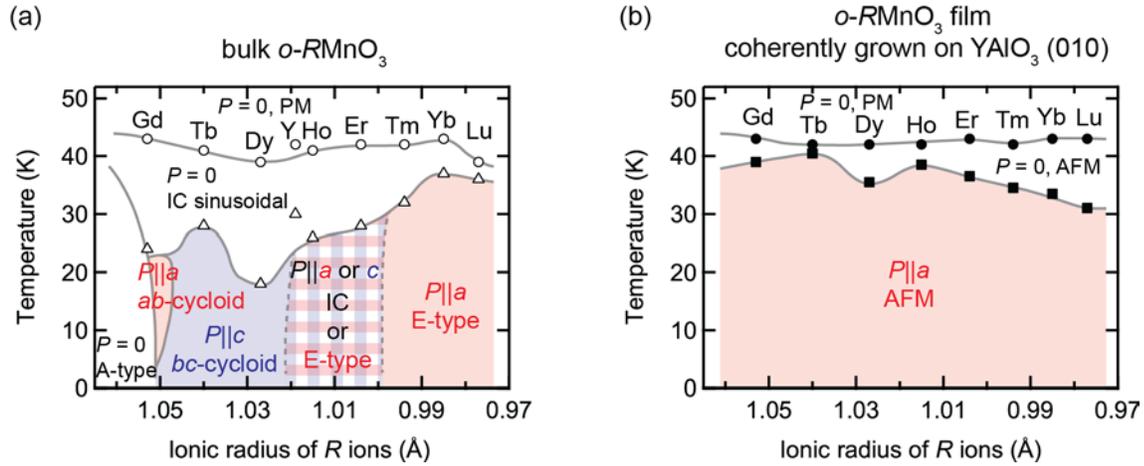

Fig. 5. (a) A multiferroic phase diagram of bulk o-$R$MnO$_3$ ($R$ = Gd – Lu, Y) based on Refs. [25,26,91]. "PM" stands for paramagnetism and "IC" for incommensurate. (b) A diagram of o-$R$MnO$_3$ ($R$ = Gd – Lu) films coherently grown on (010)-oriented YAlO$_3$ substrate. Circle and triangle markers indicate magnetic transitions and triangle and rectangle markers present FE phase transitions for each o-$R$MnO$_3$. Filled symbols are derived from the present work.

TABLE I. Magnetic and electric transition temperatures ($T_N$ and $T_{FE}$) of bulk $o$-$R$MnO$_3$ and those of films coherently grown on (010)-oriented YAlO$_3$ substrates.

| | | Gd | Tb | Dy | Ho | Er | Tm | Yb | Lu |
|---|---|---|---|---|---|---|---|---|---|
| $T_N$ (K) | Bulk | 43[a] | 41[a] | 39[a] | 41[b] | 42[c] | 42[d] | 43[e] | 39[f] |
| | Film | 43 | 42 | 42 | 42 | 43 | 43 | 43 | 43 |
| $T_{FE}$ (K) | Bulk | 8[a] (15[g])  < 30  (@ 8.4 GPa)[i] | 28[a]  ~ 33  (@ 8.7 GPa)[j] | 18[a]  ~ 31  (@ 7.1 GPa)[i] | 26[h] | 28[c] | 32[d] | 37[k] | 36[f] |
| | Film | 39 | 41[l] | 36 | 39 | 37 | 35 | 34 | 32 |

[a] Reference [23]

[b] Reference [13]

[c] Reference [67]

[d] Reference [24]

[e] Reference [66]

[f] Reference [92]

[g] Reference [93]

[h] Reference [94]

[i] Reference [71]

[j] Reference [8]

[k] Reference [25]

[l] Reference [45]

**Single-axis dependent structural and multiferroic properties of orthorhombic $R$MnO$_3$ ($R$ = Gd – Lu) – Supplementary Information –**

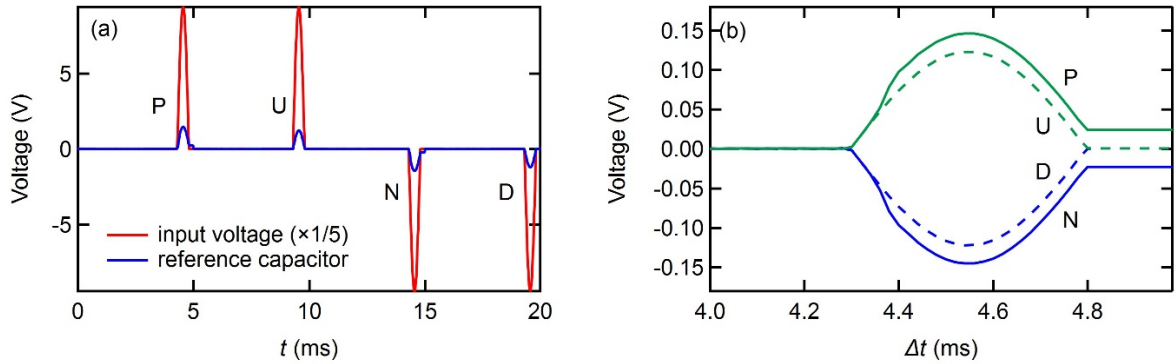

**Figure S1.** Raw data of the ferroelectric hysteresis loop measurement along the *a*-axis of the TbMnO$_3$ thin film. (a) The input voltage and the voltage measured for the reference capacitor are plotted as a function of measurement time. Prior to the first input "P," a negative voltage is applied in order to pole the ferroelectric sample. (b) Magnified plots of the voltage measured for the reference capacitor. The difference between P and U (N and D) is converted to the ferroelectric hysteresis loop at a positive (negative) electric field.